\documentclass[twocolumn]{aastex701}

\usepackage{color}
\usepackage{natbib}
\usepackage{graphicx}
\usepackage{float}
\usepackage{amsmath}
\usepackage{amssymb}
\usepackage{bm}
\usepackage[toc,page]{appendix}
\usepackage{multirow}

\shortauthors{Schroeder et al.}

\begin{document}

\author[0000-0001-9915-8147]{Genevieve~Schroeder}
\affiliation{Department of Astronomy, Cornell University, Ithaca, NY 14853, USA}\email[show]{gms279@cornell.edu}

\author[0000-0001-8405-2649]{Ben~Margalit}
\affiliation{School of Physics and Astronomy, University of Minnesota, 116 Church Street, Minneapolis, MN 55455, USA}\email{margalit@umn.edu}

\author[0000-0002-4670-7509]{Brian D.~Metzger}
\affil{Department of Physics and Columbia Astrophysics Laboratory, Columbia University, New York, NY 10027, USA}
\affil{Center for Computational Astrophysics, Flatiron Institute, 162 5th Ave, New York, NY 10010, USA}\email{bdm2129@columbia.edu}

\author[0000-0002-7374-935X]{Wen-fai Fong}
\affiliation{Center for Interdisciplinary Exploration and Research in Astrophysics (CIERA) and Department of Physics and Astronomy, Northwestern University, Evanston, IL 60208, USA}\email{wfong@northwestern.edu}

\author[0000-0002-5826-0548]{Benjamin P. Gompertz}
\affiliation{School of Physics and Astronomy, University of Birmingham, Birmingham B15 2TT, UK}
\affiliation{Institute for Gravitational Wave Astronomy, University of Birmingham, Birmingham B15 2TT, UK}\email{b.gompertz@bham.ac.uk}

\author[0000-0002-8297-2473]{Kate D.~Alexander}
\affiliation{Steward Observatory, University of Arizona, 933 North Cherry Avenue, Tucson, AZ 85721-0065, USA}\email{kdalexander@arizona.edu}

\author[0000-0002-9392-9681]{Edo Berger}
\affiliation{Center for Astrophysics | Harvard \& Smithsonian, Cambridge, MA 02138, USA}\email{eberger@cfa.harvard.edu}

\author[0000-0003-1792-2338]{Tanmoy Laskar}
\affiliation{Department of Physics \& Astronomy, University of Utah, Salt Lake City, UT 84112, USA}
\affiliation{Department of Astrophysics/IMAPP, Radboud University, P.O. Box 9010, 6500 GL, Nijmegen, The Netherlands}\email{tanmoy.laskar@utah.edu}

\author[0000-0001-5169-4143]{Gavin P. Lamb}
\affiliation{Astrophysics Research Institute, Liverpool John Moores University, IC2 Liverpool Science Park, 146 Brownlow Hill, Liverpool, L3 5RF, UK}\email{g.p.lamb@ljmu.ac.uk}

\author[0000-0001-7821-9369]{Andrew Levan}\affiliation{Department of Astrophysics/IMAPP, Radboud University, P.O. Box 9010, 6500 GL Nijmegen, The Netherlands}\email{a.levan@astro.ru.nl}

\author[0000-0002-5740-7747]{Charles~D.~Kilpatrick}
\affiliation{Center for Interdisciplinary Exploration and Research in Astrophysics (CIERA) and Department of Physics and Astronomy, Northwestern University, Evanston, IL 60208, USA}\email{ckilpatrick@northwestern.edu}

\author[0000-0002-9267-6213]{Jillian C.~Rastinejad}
\altaffiliation{NHFP Einstein Fellow}
\affiliation{Department of Astronomy, University of Maryland, College Park, MD 20742, USA}\email{jcrastin@umd.edu}

\title{No Sign of a Magnetar Remnant Following the Kilonova-Producing Long GRB 211211A $\sim 1.7~$Years Later}

\begin{abstract}
    In addition to a $\gamma$-ray burst (GRB), the merger of two neutron stars may produce a temporarily or indefinitely stable neutron star remnant with a strong magnetic field (a ``magnetar''). As this magnetar remnant spins down, it can deposit its rotational energy into the surrounding kilonova ejecta, producing synchrotron emission that peaks in the radio bands $\sim$months--years after the merger (``boosted kilonova''). The nearby ($z=0.0763$) long-duration GRB\,211211A, which has an apparent kilonova counterpart and likely neutron star merger progenitor, may have produced such a remnant.
    We observed the location of GRB\,211211A at 6~GHz with the NSF's Karl G. Jansky Very Large Array (VLA) spanning $\approx 0.54$--$1.7~$years after the burst. We do not detect any radio emission, placing strong limits on the energy deposited into the ejecta by any remnant to $\lesssim 4.4 \times 10^{52}~{\rm erg}$. Due to the proximity of the event, we are also able to place limits on a kilonova afterglow that did not receive any additional energy deposition, though it is possible such emission will be suppressed until $\sim 4~{\rm years}$ after the burst, when the kilonova is expected to overtake the forward shock of the GRB. Future observations with the VLA and next-generation radio facilities will be able to further constrain the magnetar-boosted kilonova and kilonova afterglow scenarios, as well as directly constrain models in the scenario that GRB\,211211A was instead produced by a collapsar.
\end{abstract}

\section{Introduction}

The merger of neutron stars (NSs) produces gravitational wave (GW) emission as well as a $\gamma$-ray burst (GRB), as demonstrated by the multimessenger detection of GW\,170817/GRB\,170817 \citep{ 2017ApJ...848L..13A, 2017ApJ...848L..12A}. Such mergers are expected to produce neutron-rich ejecta by tidal stripping and accretion disk outflows, which synthesize radioactive nuclei that power a thermal red transient (``kilonova'', \citealt{2010MNRAS.406.2650M}). This kilonova signal was observed following GW\,170817 \citep[AT2017gfo, e.g.][]{coulter17,chornock17,pian17}, confirming that NS mergers produce some of the heaviest elements in the universe via rapid neutron capture nucleosynthesis \citep[``$r$-process'', for a review see e.g.][]{2021ARA&A..59..155M}. In addition to NS mergers, GRBs can be produced by massive stellar explosions (``collapsars''), and this progenitor scenario has been confirmed with the routine detection of stripped-envelope supernovae following GRBs \citep[e.g.][]{1998Natur.395..670G, 1999ApJ...524..262M, 2006ARA&A..44..507W,2006Natur.441..463F,2017AdAst2017E...5C, 2019NatAs...3..717M}.
It is typically assumed (and in some cases, proven) that the GRBs with $\gamma$-ray emission lasting $\lesssim 2~{\rm s}$ are produced by NS mergers, while those with $\gamma$-ray emission lasting $\gtrsim 2~{\rm s}$ originate from collapsars  \citep[``short'' and ``long'' GRBs, respectively,][]{1993ApJ...413L.101K}. However, there have been a few examples of long-duration GRBs that are more consistent with an NS merger progenitor rather than a collapsar progenitor \citep{2006Natur.444.1050D, 2006Natur.444.1047F, 2006Natur.444.1053G,  2006Natur.444.1044G, 2022Natur.612..223R, 2022Natur.612..228T, 2022Natur.612..232Y, 2024Natur.626..737L, 2024Natur.626..742Y, 2025NSRev..12E.401S}. \citet{2025ApJ...979..190R} uniformly modeled a sample of GRBs (both traditionally short GRBs and three long GRBs) with claimed kilonovae, and found that the long GRBs with kilonovae may have higher inferred kilonova ejecta masses, potentially indicating asymmetric mass ratios for NS mergers that produce long GRBs (however see \citealt{2025ApJ...982...97K} for a detailed investigation of model inferred masses for kilonovae). Despite this possible difference, their afterglow luminosities remain unremarkable compared to the traditional short GRB population, further substantiating the merger progenitor scenario \citep{2025arXiv250820156C}.

One notable example that has garnered high interest is GRB\,211211A, a nearby ($z = 0.0763$) GRB with a long duration \citep[$\sim 30$--$50~{\rm s}$,][]{2021GCN.31209....1S, 2021GCN.31210....1M}, yet has no associated supernova to optical luminosities of $\lesssim 3 \times 10^{40}~{\rm erg~s}^{-1}$, $\sim 200$ times fainter than the prototypical GRB-SN1998bw \citep{2022Natur.612..223R}. Instead, the resulting optical and infrared emission revealed a red transient consistent with the color and temporal behavior of AT2017gfo, leading to its identification as a kilonova, and a general consensus that GRB\,211211A was likely produced by an compact object merger
{\citep{2022Natur.612..223R, 2022Natur.612..228T, 2022Natur.612..232Y,2022A&A...664A.177S, 2023ApJ...947L..21Z, 2024MNRAS.527.3900K, 2024MNRAS.527.7722S, 2024ApJ...963..112M, 2025ApJ...988L..46L, 2025MNRAS.540.2727L,Cheong+25}}. However, whether the merger was between two NSs, a white dwarf (WD) and NS, two WDs, or a black hole (BH) and NS, is still undetermined. While some studies have suggested a NS merger with a WD \citep[e.g.][]{2022Natur.612..232Y}, most simulations of WD-NS mergers find insufficient accretion rates to generate neutron-rich disk outflows and $r$-process elements (e.g., \citealt{Margalit&Metzger16,Fernandez+19}), and such systems may, in fact, be rarer than NS-NS mergers \citep{chrimes25}.
Additionally, a collapsar origin, in which the red color comes from either thermal dust emission (at a higher redshift) or an unusual $r$-process rich supernova, has also been proposed to explain the red transient \citep{2023ApJ...947...55B, 2025ApJ...984...37W, 2025arXiv250903003R}, although the host offset of the transient disfavor these explanations \citep{2022Natur.612..223R}.  \added{While a low-mass collapsar which produces a GRB through fallback accretion onto a BH, rather than through direct collapse, could explain the lack of a luminous supernova \citep[e.g.][]{2007ApJ...662L..55F}, it is difficult to reconcile the infrared excess observed following GRB\,211211A in this scenario \citep{2022Natur.612..228T}, unless some $r$-process material is also produced \citep{2023ApJ...947...55B, 2025arXiv250903003R}.} In the merger scenario (involving NSs or WDs), one of the proposed explanations for the prolonged $\gamma$-ray duration is that the merger produced a highly magnetized NS remnant {\citep[``magnetar'';][]{2022Natur.612..223R, 2022A&A...664A.177S, 2022Natur.612..232Y, 2023ApJ...947L..21Z,Cheong+25,Combi+25}}. In this scenario, the proto-magnetar created in the merger will produce a relativistic wind which then powers a longer emission episode \citep{2008MNRAS.385.1455M, 2011MNRAS.413.2031M, 2012MNRAS.419.1537B, 2013MNRAS.431.1745G, 2014MNRAS.438..240G}. Since magnetar spin-down parameters are set by the magnetic field and initial spin, the extraction of energy from a magnetar naturally occurs on timescales much longer than the dynamical timescale of the merger. One way of testing these different progenitor scenarios, as well as diagnosing the cause of the long duration, is to constrain the presence of a merger remnant, and determine whether a magnetar was produced \citep{2025MNRAS.539.1908O}.

The nature of the merger remnant (i.e., a NS that survives for any length of time versus a BH) is dependent on the total masses of the merger components, as well as the maximum mass of a cold non-rotating NS (Tolman–Oppenheimer–Volkoff mass, $M_{\rm TOV}$). This maximum mass is not well constrained due to the unknown NS equation of state (EoS). However, observations of pulsars, short GRBs, and GW\,170817 have placed constraints on the range of plausible $M_{\rm TOV}$ \citep[e.g. ][]{2015ApJ...808..186L, 2017ApJ...850L..19M, 2019PhRvD.100b3015S, 2020NatAs...4...72C}. If the mass of the merger remnant greatly exceeds $M_{\rm TOV}$, then it is expected to collapse directly into a BH \citep{2013PhRvL.111m1101B}. However, for somewhat less massive remnants $(M \lesssim 1.2$--$1.3 ~M_{\rm TOV}$), rotation is able to prevent immediate gravitational collapse \citep{Margalit2022} and the merger remnant may also exist as a temporarily stable magnetar that collapses to a BH in milliseconds (so-called hypermassive NSs), or for seconds to minutes (supermassive NSs). Remnants with the lowest masses ($M < M_{\rm TOV}$) may remain as stable magnetars even after they spin down \citep{2013ApJ...771L..26G}.  These different outcomes make drastically different predictions regarding the energy released into the environment.

Assuming a magnetar remnant is produced following an NS merger, and its outflow is effectively coupled to its environment, the lost rotational energy \citep[up to $\sim 10^{53}~{\rm erg}$ for $M_{\rm TOV} \sim 2.2~M_\odot$;][]{2015MNRAS.454.3311M} will be deposited into the surrounding kilonova ejecta \citep{Yu+13,Metzger&Piro14,Fang&Metzger17,Sarin+22,Ai+25}. As a result, this ``magnetar-boosted kilonova'' is expected to produce synchrotron emission as it interacts with the surrounding interstellar medium (ISM), with this emission peaking at $\sim{\rm GHz}$ frequencies $\sim{\rm years}$ after the merger \citep{np2011Natur.478...82N, 2014MNRAS.437.1821M,2015MNRAS.450.1430H, 2025MNRAS.539.1908O}. Detecting such a radio transient would not only confirm the production of a magnetar remnant, but also would have implications for the NS EoS. 

Several radio searches for a magnetar-boosted kilonova have been performed following dozens of GRBs produced by NS mergers. However, there are no confirmed detections thus far \citep{2014MNRAS.437.1821M, 2016ApJ...831..141F, 2016ApJ...819L..22H, 2019ApJ...887..206K, 2020ApJ...890..102L, smf+2020ApJ...902...82S,2021MNRAS.505L..41B, 2021MNRAS.500.1708R, 2021ApJ...908...63G, 2021ApJ...923...38N, 2021A&A...650A.117N, 2023ApJ...948..125E, 2024MNRAS.527.8068G}. Many of these studies are only able to place weak limits on the energy imparted onto the kilonova by a potential magnetar remnant ($\lesssim 10^{53}~$erg) due to the typically large distances to these mergers \citep[$z \lesssim  3$, with a median of $\langle z \rangle \approx 0.6$, ][]{2022MNRAS.515.4890O, 2022ApJ...940...56F, 2022ApJ...940...57N}. As a result, it is difficult to constrain the presence of a temporarily stable magnetar remnant for these events, though strong limits of $\lesssim 1.3 \times 10^{52}~$erg have been placed for the nearby merger GW\,170817 \citep{smf+2020ApJ...902...82S}.

Even if a merger does not produce a magnetar remnant, the kilonova ejecta has intrinsic kinetic energy ($\sim 10^{51}~$erg), and in turn may produce a radio signature as the kilonova interacts with the surrounding medium
\citep[``kilonova afterglow'',][]{np2011Natur.478...82N, 2013MNRAS.430.2121P, 2014MNRAS.437L...6K, 2014PhRvD..89f3006T, 2015MNRAS.450.1430H, 2018ApJ...867...95H, 2019MNRAS.487.3914K, 2020MNRAS.495.4981M}. However, the kilonova afterglow is expected to be much fainter than the magnetar-boosted kilonova emission, and therefore limits on this emission can only be placed on very nearby NS mergers \citep[$z \lesssim 0.1$, ][]{np2011Natur.478...82N}, such as GW\,170817 \citep{2019ApJ...886L..17H, 2022ApJ...927L..17H}.

Given its proximity ($\sim 350~{\rm Mpc}$), and its unusual $\gamma$-ray properties, GRB\,211211A provides one of the best opportunities to search for radio emission from a magnetar-boosted kilonova, as well as the kilonova afterglow. Such observations offer unique insights into the various proposed progenitor models of merger-driven long GRBs.
Here we present observations from the NSF's Karl G. Jansky Very Large Array (VLA) of GRB\,211211A, to place constraints on two radio emission scenarios: a magnetar-boosted kilonova and kilonova afterglow. In Section~\ref{sec:Observations} we present the details of our radio observations of GRB\,211211A taken at $\sim 0.54$--$1.7~$yr after the burst. In Section~\ref{sec:Methods} we present our updated light curve modeling, which we use in Section~\ref{sec:Results} to explore the parameter space that our radio observations can rule out for both radio emission scenarios. We discuss the implications of our observations for the various proposed progenitors of GRB\,211211A, as well as future observational prospects, in Section~\ref{sec:Discussion}. We conclude in Section~\ref{sec:Conclusions}. In this paper, we employ the $\Lambda$CDM cosmological parameters of $H_{0} = 68~{\rm km \, s}^{-1} \, {\rm Mpc}^{-1}$, $\Omega_{M} = 0.31$, $\Omega_{\rm \Lambda} = 0.69$ \citep{2020A&A...641A...6P}.

\begin{deluxetable}{ccc}[!t]
 \tabletypesize{\normalsize}
 \tablecolumns{3}
 \tablecaption{VLA Radio Observations}
 \tablehead{ 
   \colhead{Observation Date} &
   \colhead{$\delta t^{\rm{a}}$} &
   \colhead{$F_\nu^{b}$} \\
   \colhead{(UT)} &
   \colhead{(days)} &
   \colhead{($\mu$Jy)}
   }
\startdata 
    2022-06-26 & 197.4 	& $< 10.2$ \\
    2022-08-27 & 259.3 	& $< 13.2$ \\
    2022-12-23 & 377.0 	& $< 6.9$ \\
    2023-08-11 & 608.4 	& $< 8.4$ \\
\enddata
\tablecomments{$^{\rm{a}}$ Mid-time of entire observation compared to {\it Swift} trigger.\\
$^{b}$ Upper limits correspond to $3 \sigma$ confidence.
}
\label{tab:radio_data}
\end{deluxetable}

\section{Observations}
\label{sec:Observations}

We were awarded Director's Discretionary time to observe the position of GRB\,211211A with the Karl G. Jansky Very Large Array (VLA) at a central frequency of 6\,GHz (4.04--7.96~GHz). We obtained three epochs at $\delta t \approx 197$--$377$~days (where $\delta t$ refers to the time since the {\it Swift} trigger) under program 22A-495 (PI~Schroeder), and an additional epoch at $\delta t \approx 608~$days under program 23A-298 (PI~Schroeder). We used 3C286 for band-pass and flux calibration and J1407+2827 for gain and phase calibration. We reduced the data using the Common Astronomy Software Applications Pipeline \citep[\texttt{CASA}, ][]{CASA}, performing manual flagging when necessary, and image using standard routines. We do not detect any emission at or around the location of GRB\,211211A in any of our epochs. In order to measure the RMS of the image, we use the \texttt{pwkit/imtool} program \citep{2017ascl.soft04001W}, and place $3\sigma$ limits of $\lesssim 7-13 ~\mu$Jy at the location of GRB\,211211A at $\delta t \approx 0.54$--$1.7~{\rm years}$. A summary of the observations can be found in Table~\ref{tab:radio_data}.

\begin{figure*}
    \centering
    \includegraphics[width = 0.9\textwidth]{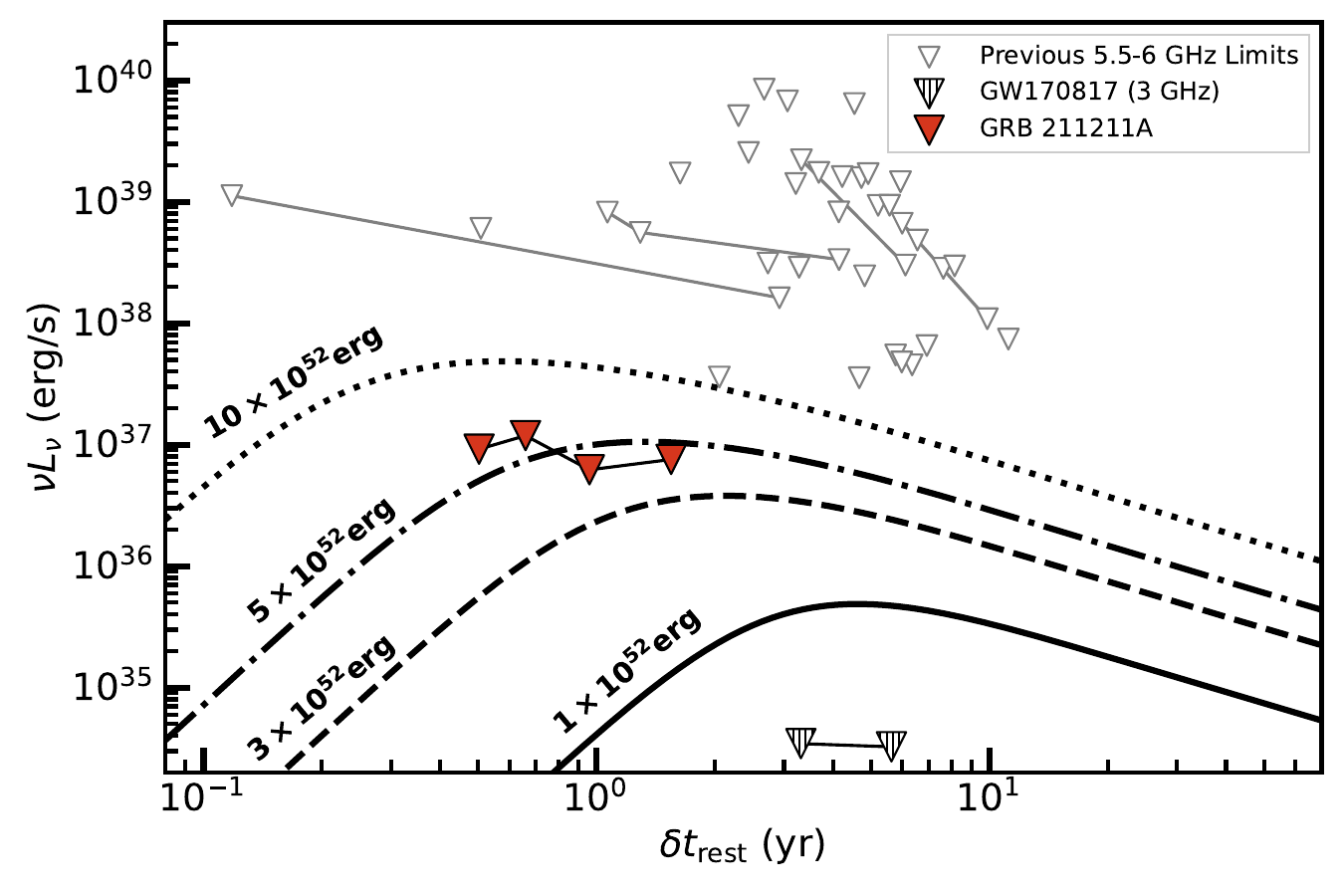}
    \caption{Radio luminosity $\nu L_{\nu}$ vs. rest frame time since gamma-ray trigger ($\delta t_{\rm rest})$. Orange triangles represent $3\sigma$ limits of our 6\,GHz observations of GRB\,211211A. Empty triangles represent previous 6\,GHz limits of short GRBs \citep{2016ApJ...831..141F, 2019ApJ...887..206K, smf+2020ApJ...902...82S, 2021ApJ...923...38N, 2021MNRAS.500.1708R, 2021MNRAS.505L..41B, 2021ApJ...923...38N, 2021A&A...650A.117N}, whereas hatched triangles represent the deepest 3~GHz limits for GW\,170817 \citep{2022ApJ...927L..17H, 2024GCN.36390....1E}. Black lines represent single velocity shell light-curve models for a variety of ejecta energies ($E_{\rm ej} = 1,3,5,10 \times 10^{52}~$erg). We assume the afterglow-derived values for $p$, $n_0$, $\epsilon_{\rm e}$, $\epsilon_{\rm B}$, and the kilonova-derived value for $M_{\rm ej}$ (Table~\ref{tab:model}). Given that GRB\,211211A was relatively nearby, the non-detections place strong luminosity limits on any radio emission.}
    \label{fig:LightCurves}
\end{figure*}

\section{magnetar-boosted Kilonova and Kilonova Afterglow Models}
\label{sec:Methods}

We create a suite of model light-curves using a revised version of the model described in \cite{smf+2020ApJ...902...82S}. We consider two ejecta shell scenarios: a `single velocity shell' scenario and a `structured ejecta' scenario. We use these model light curves to translate our VLA non-detections into constraints on various parameters in both magnetar-boosted kilonova (Section~\ref{sec:Magnetar_Constraints})
and ordinary kilonova afterglow (Section~\ref{sec:KN_afterglow}) scenarios for GRB\,211211A.

For the single velocity shell scenario, we calculate the dynamics of ejecta-ISM forward shock following \cite{2012ApJ...752L...8P} as in \cite{smf+2020ApJ...902...82S}. The key model parameters are the energy deposited into the kilonova ejecta ($E_{\rm ej}$), the mass of the kilonova ejecta ($M_{\rm ej}$), the density of the circumburst material ($n_0$), the non-thermal electron power law distribution index ($p$), and the fractions of energy imparted on the electrons and magnetic field ($\epsilon_{\rm e}$ and $\epsilon_{\rm B}$, respectively). The ejecta velocity is derived from $M_{\rm ej}$ and $E_{\rm ej}$.

For the structured ejecta scenario, the kilonova ejecta is considered to be radially stratified such that the kinetic energy of ejecta moving at four-velocities greater than $\Gamma \beta$ is $E_{\rm ej}(\geq \Gamma\beta) \propto (\Gamma \beta)^{-\alpha}$, where $\beta$ is the velocity of the kilonova ejecta, $\Gamma$ is the Lorentz factor of the ejecta, and $\alpha$ is the power-law distribution of the energy, as described in e.g. \cite{2019MNRAS.487.3914K}.
The dynamics in this scenario are analytically calculated following the formalism of \cite{2015MNRAS.450.1430H} (their Eq.~9).
The synchrotron emission is calculated following the formalism of \cite{2021ApJ...923L..14M}, which accounts for the effects of synchrotron self-absorption, synchrotron cooling, and the impact of the post-shock thermal electrons on the resulting emission. 
We use a revised version of this formalism that accounts for additional relativistic effects which were not included in the original \cite{2021ApJ...923L..14M} paper, and become important when $\Gamma\beta \gtrsim 1$ \citep{2024ApJ...977..134M}. 
Note that this analytic approach treats relativistic effects following a line-of-sight approximation, similar to e.g. \citet{1998ApJ...497L..17S}; though see \citet{2025arXiv250916313F} for a full numerical treatment of relativistic effects. 
In our present work, we choose to neglect thermal electrons in the light-curve modeling. This approach is chosen primarily for the sake of simplicity (among other things, it reduces the number of free parameters) and in order to facilitate comparison with previous work where thermal electrons were ignored. Furthermore, the absence or presence of thermal electrons in mildly-relativistic transients is still actively debated \citep[e.g.,][]{2022ApJ...932..116H,2025arXiv250613618R,2025arXiv250900952N}, so our present choice reverts to default conventions in the literature.
The key parameters of the structured ejecta scenario are the same as the single velocity shell scenario, with the addition of $\alpha$. 

\section{Results}
\label{sec:Results}
\begin{deluxetable}{ccc}[!t]
 \tabletypesize{\normalsize}
 \tablecolumns{2}
 \tablecaption{Model parameters}
 \tablehead{ 
   \colhead{Parameter} &
   \colhead{Afterglow/Kilonova Derived$^a$} &
   \colhead{Fiducial}
   }
\startdata 
    $p$ 	& $2.014^{+0.007}_{-0.003}$ & $2.2$ \\
    $n_{0}~({\rm cm}^{-3})$ 	& $0.54^{+10.02}_{-0.52}$ & $0.54^{+10.02}_{-0.52}$\\
    $\epsilon_{\rm e}$ 	& $3.28^{+14.8}_{-2.77}\times 10^{-2}$ & $0.1$ \\
    $\epsilon_{\rm B}$ 	& $1.52^{+50.5}_{-1.44} \times 10^{-4}$ & $0.01$ \\
    $M_{\rm ej}~(M_\odot)$ 	& $0.047^{+0.026}_{-0.011}$ & $0.047^{+0.026}_{-0.011}$\\
    \hline
    $E_{\rm ej, max}~({\rm erg})$ 	& $4.4 \times 10^{52}$ & $6.1 \times 10^{51}$\\
\enddata
\tablecomments{$^a$Values from \citet{2022Natur.612..223R}
}
\label{tab:model}
\end{deluxetable}

\begin{figure*}
    \centering
    \includegraphics[width = 0.9\textwidth]{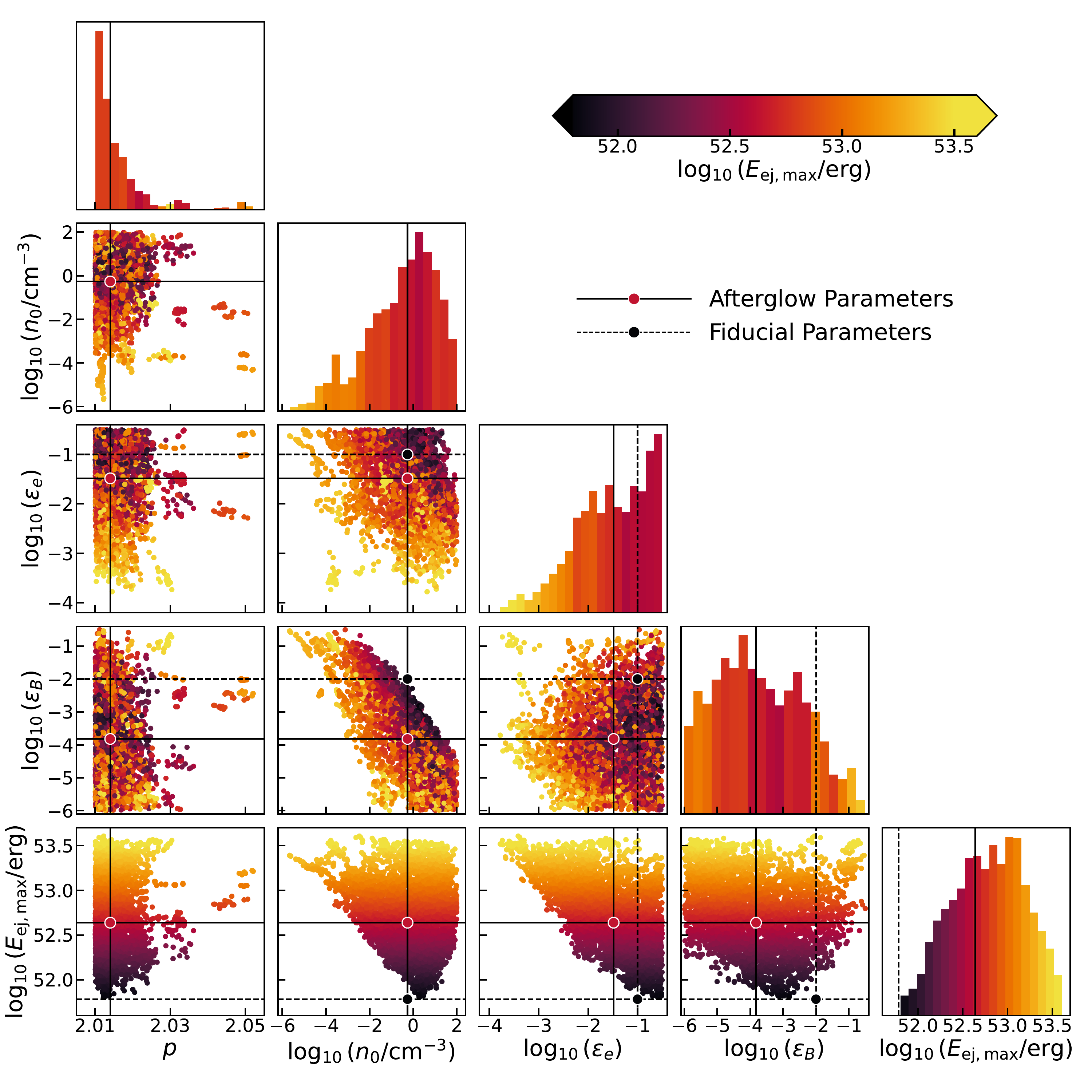}
    \caption{Corner plot showing the one and two dimensional projections of the 5300 samples of the posterior distributions of the afterglow model presented in \citet{2022Natur.612..223R}. Points are colored by the maximum ejected energy ($E_{\rm ej, max}$) consistent with our radio limits for a given parameter sample. The posterior density functions for each parameter are split into 20 bins, and each bin is colored by the median $E_{\rm ej, max}$ for the samples that fall within a particular bin. The bottom row depicts the dependence on $E_{\rm ej, max}$ with each parameter. The solid (dashed) black line and red (black) point denote the solution assuming the afterglow (fiducial) parameters, with the color corresponding to $E_{\rm ej, max} \approx 4.4 \times 10^{52}~$erg ($\approx 6.1 \times 10^{51}~$erg).}
    \label{fig:ParameterSpace}
\end{figure*}

Here, we utilize the radio observations of GRB\,211211A (Section~\ref{sec:Observations}) to compare to the generated light curve models described in Section~\ref{sec:Methods} in order to place limits on the presence of magnetar-boosted kilonova emission following GRB\,211211A, as well as place limits on a kilonova afterglow. For direct comparisons to previous works \citep[e.g.][]{2016ApJ...831..141F, 2016ApJ...819L..22H, 2019ApJ...887..206K, 2020ApJ...890..102L, smf+2020ApJ...902...82S, 2021MNRAS.500.1708R, 2021MNRAS.505L..41B, 2024MNRAS.527.8068G}, we derive the limits we can place on $E_{\rm ej}$ in the magnetar-boosted kilonova scenario assuming the single velocity shell model. Similarly, we assume the structured ejecta model when considering the kilonova afterglow scenario to directly compare to previous studies on GW\,170817A \citep[e.g.][]{2019ApJ...886L..17H, 2022ApJ...927L..17H}.

\subsection{Constraints on Magnetar Production}
\label{sec:Magnetar_Constraints}

\subsubsection{Preliminary Constraints}
\label{sec:Preliminary_Constraints}

Given the low redshift of GRB\,211211A, our observations are able to probe much lower luminosities ($\nu L_{\nu} \lesssim 10^{37}~$erg s$^{-1}$, Figure~\ref{fig:LightCurves}) than previous radio studies of short GRBs at similar observing frequencies \citep[$\gtrsim 4 \times 10^{37}~$erg/s, ][]{2016ApJ...831..141F, 2019ApJ...887..206K, smf+2020ApJ...902...82S, 2021ApJ...923...38N, 2021MNRAS.500.1708R, 2021MNRAS.505L..41B, 2021ApJ...923...38N, 2021A&A...650A.117N}. Motivated by these deep luminosity limits, we explore the constraints that our observations can place on $E_{\rm ej}$ for the magnetar-boosted kilonova scenario using the single shell velocity model. For the remaining model parameters, we consider two sets of values, listed in Table~\ref{tab:model}. First, we use the median values inferred from the afterglow for $p$, $n_0$, $\epsilon_{\rm e}$, and $\epsilon_{\rm B}$, as well as the derived $M_{\rm ej}$ from the kilonova in \citet{2022Natur.612..223R}. The summary of the assumed parameters for this scenario are provided in the first column of Table~\ref{tab:model}. We generate a suite of light curves for $E_{\rm ej} = 10^{52}$--$10^{53}~$erg in Figure~\ref{fig:LightCurves} using the afterglow parameters. Overall, these values are similar to those derived in other studies of GRB\,211211A \citep[e.g.][]{2022Natur.612..228T, 2022Natur.612..232Y}, except for $p$, which we assume to be $p\approx 2.014$, whereas other studies find $p \approx 2.1$--$2.2$. We explore how our choices of parameters affect our results below.

From a comparison to the single velocity shell models, we determine the maximum ejecta energy ($E_{\rm ej, max}$) allowed by our non-detections to be $E_{\rm ej, max} \approx 4.4\times 10^{52}~$erg (Figure~\ref{fig:LightCurves}). Given that the maximum energy for an indefinitely stable magnetar is $E_{\rm ej} \sim 10^{53}~$erg, this limit indicates that if GRB 2112111A produced a magnetar remnant, it was not indefinitely stable. 

However, it is not clear if the microphysical parameters ($p$, $\epsilon_{\rm e}$, and $\epsilon_{\rm B}$) of the afterglow and magnetar-boosted kilonova should be the same. To investigate the effects of this assumption, we also consider a second set of parameters using {\it fiducial} values drawn from the general GRB population \citep[e.g.][]{2015ApJ...815..102F, 2023ApJ...959...13R} of $p=2.2$, $\epsilon_{\rm e}=0.1$, and $\epsilon_{\rm B}=0.01$ (using the same values as derived from the afterglow or kilonova for all other parameters; second column of Table~\ref{tab:model}). From the comparison of the observations to the suite of light curves, we find a more stringent $E_{\rm ej, max} \approx 6.1\times 10^{51}~$erg, which effectively rules out a supramassive, temporarily stable magnetar remnant \citep[$E_{\rm ej} \sim 10^{52}~{\rm erg}$,][]{2019ApJ...880L..15M}. We note that our fiducial parameters result in a much more stringent limit on $E_{\rm ej, max}$ primarily because of the higher assumed value of $p$ compared to that directly inferred from the afterglow. This is due to our light curve function scaling\footnote{We note that this approximation breaks downs as $p \to 2$. However, our constraint on $E_{\rm ej, max}$ for $p = 2.014$ is more conservative than the constraint that would be calculated using a more comprehensive treatment of this effect.}
as $F_{\nu}\propto (p-2)$, resulting in low luminosities when $p\sim 2$ ($p = 2.014$ for the afterglow parameters, compared to $p = 2.2$ for the fiducial parameters). Compared to the analysis of 27 short GRBs using similar methods presented in \citet{smf+2020ApJ...902...82S}, our limit of $E_{\rm ej, max} \approx 4.4 \times 10^{52}~{\rm erg}$ is deeper than the limits placed on $\sim 40-70\%$ of cosmological short GRBs\footnote{If we assume  $E_{\rm ej, max} \approx 6.1 \times 10^{51}~{\rm erg}$ only one short GRB has had deeper limits placed \citep{smf+2020ApJ...902...82S}.}.

\subsubsection{Distribution of $E_{\rm ej, max}$}

We next explore the relationship between $E_{\rm ej, max}$ and the parameters $p$, $n_0$, $\epsilon_{\rm e}$, and $\epsilon_{\rm B}$. In Section~\ref{sec:Preliminary_Constraints}, we assumed that the magnetar-boosted kilonova inherits the same parameter values as the afterglow. However, for preliminary constraints on $E_{\rm ej, max}$, we only used the median values of each parameter, while the full posterior distributions for some parameters span several orders of magnitude (Table~\ref{tab:model}, \citealt{2022Natur.612..223R}). These large uncertainties can affect the resulting conclusion on magnetar production. Thus, we leverage the full posterior distributions of $p$, $n_0$, $\epsilon_{\rm e}$, and $\epsilon_{\rm B}$ from the afterglow model fitting \citep{2022Natur.612..223R}, to compute the distribution of $E_{\rm ej, max}$. We also investigate any trends between the afterglow parameters and $E_{\rm ej, max}$.

We start by sampling the posterior distributions of the afterglow model presented in \citet{2022Natur.612..223R} and extract 5300 samples (where one ``sample'' corresponds to a set of linked values for $p$, $n_0$, $\epsilon_{\rm e}$, and $\epsilon_{\rm B}$). We then produce the corresponding model light curves for a range of $E_{\rm ej}$ ($E_{\rm ej} = 10^{50}$--$10^{54}~$erg), and determine for each sample the largest value $E_{\rm ej, max}$ that does not violate our non-detections. The full distributions are displayed in Figure~\ref{fig:ParameterSpace}, and are colored by $E_{\rm ej, max}$. For comparison, we also denote the values of the afterglow and fiducial parameters assumed in Section~\ref{sec:Preliminary_Constraints}.

We first explore what the distribution of $E_{\rm ej, max}$ indicates for the remnant of GRB\,211211A. The median $E_{\rm ej, max}$ for the 5300 samples is $\langle E_{\rm ej,max}\rangle \approx 7.4 \times 10^{52}~$erg, within a factor of $\sim 2$ of the $E_{\rm ej, max}$ derived from the afterglow parameters in Section~\ref{sec:Preliminary_Constraints}.
We find that $\sim 66\%$ of samples result in $E_{\rm ej, max}  < 10^{53}~$ergs, indicating that a stable magnetar was likely not formed by GRB\,211211A. However, only $\sim 2\%$ of samples result in  $E_{\rm ej, max}  < 10^{52}~$ergs, indicating that we cannot rule out a temporarily stable magnetar.

We next explore which parameters affect the limits placed on $E_{\rm ej, max}$ the most.
Past studies have attempted to constrain some combination of the many parameters involved in the magnetar-boosted kilonova model; typically $E_{\rm ej, max}$, $n_0$, and $\epsilon_{\rm B}$ \citep[e.g.][]{2016ApJ...831..141F, 2016ApJ...819L..22H, smf+2020ApJ...902...82S, 2020ApJ...890..102L, 2024MNRAS.527.8068G}, while holding all other parameters fixed. In our light curves, $F_{\nu} \propto n_0^{(p+1)/4}$, and therefore the naive expectation is that high values of $n_0$ would result in the strongest limits on $E_{\rm ej, max}$. However, if it is assumed that the magnetar-boosted kilonova parameters are the same as the afterglow parameters, the inherent relationships between the afterglow parameters must be taken into account.

In contrast to the expectations, we find that the expected dependence of $E_{\rm ej, max}$ on $n_0$ appears to be countered with the negative relationship between $n_0$ and $\epsilon_{\rm B}$ within the synchrotron afterglow model.
This results in high values of $E_{\rm ej, max}$ for high-$n_0$/low-$\epsilon_{\rm B}$ (and low-$n_0$/high-$\epsilon_{\rm B}$) pairs (Figure~\ref{fig:ParameterSpace}), with $n_0$/$\epsilon_{\rm B}$ pairs in the middle of their distributions leading to the strongest constraints on $E_{\rm ej, max}$.
We instead note that $E_{\rm ej, max}$ appears to have the strongest dependence on $\epsilon_{\rm e}$, where higher $\epsilon_{\rm e}$ leads to lower $E_{\rm ej, max}$. Historically, $\epsilon_{\rm e}$ is fixed to $0.1$ when assessing the limits that can be placed on $E_{\rm ej, max}$ in the magnetar-boosted kilonova scenario \citep{2016ApJ...831..141F, 2016ApJ...819L..22H, 2019ApJ...887..206K, 2020ApJ...890..102L, smf+2020ApJ...902...82S, 2021MNRAS.500.1708R, 2021MNRAS.505L..41B, 2024MNRAS.527.8068G}; however this test indicates that this may be an overly simplistic assumption.

Overall, this exploration demonstrates the sensitivity of $E_{\rm ej, max}$ on the assumed parameters. We find that our limit on $E_{\rm ej, max}$ is more stringent if we assume fiducial parameters ($\sim 6.1 \times 10^{51}~$erg, Section~\ref{sec:Preliminary_Constraints}) for the magnetar-boosted kilonova, and less stringent if we utilize the full posterior distribution of the afterglow parameters ($\sim 7.4\times 10^{52}~$erg). Given the uncertainty in the magnetar-boosted parameters, we proceed with the limit derived from the median afterglow/kilonova derived parameters in Table~\ref{tab:model}, $E_{\rm ej, max} \approx4.4\times 10^{52}~$erg (Section~\ref{sec:Preliminary_Constraints}).
{We note that the quantity we are directly constraining here is $E_{\rm ej, max}$, 
and that this is roughly the initial magnetar rotational energy only if the spin-down energy is strongly coupled to the ejecta. This may not necessarily be the case if there is significant gravitational-wave spin-down (due to deformations caused by secular instabilities or strong internal magnetic fields) or if an appreciable amount of energy is channeled into an off-axis magnetized jet or outflow (\citealt{1995ApJ...442..259L, 2005ApJ...634L.165S, 2009MNRAS.398.1869D, 2015ApJ...798...25D}). However, in physically-realizable scenarios, we expect that these effects do not play an important role or alter our overall conclusions on the nature of the merger remnant (see \citealt{2017ApJ...850L..19M} for a more thorough exploration). }

\subsection{Constraints on Kilonova Afterglow}
\label{sec:KN_afterglow}
\begin{figure*}
    \centering
    \includegraphics[width = 0.9\textwidth]{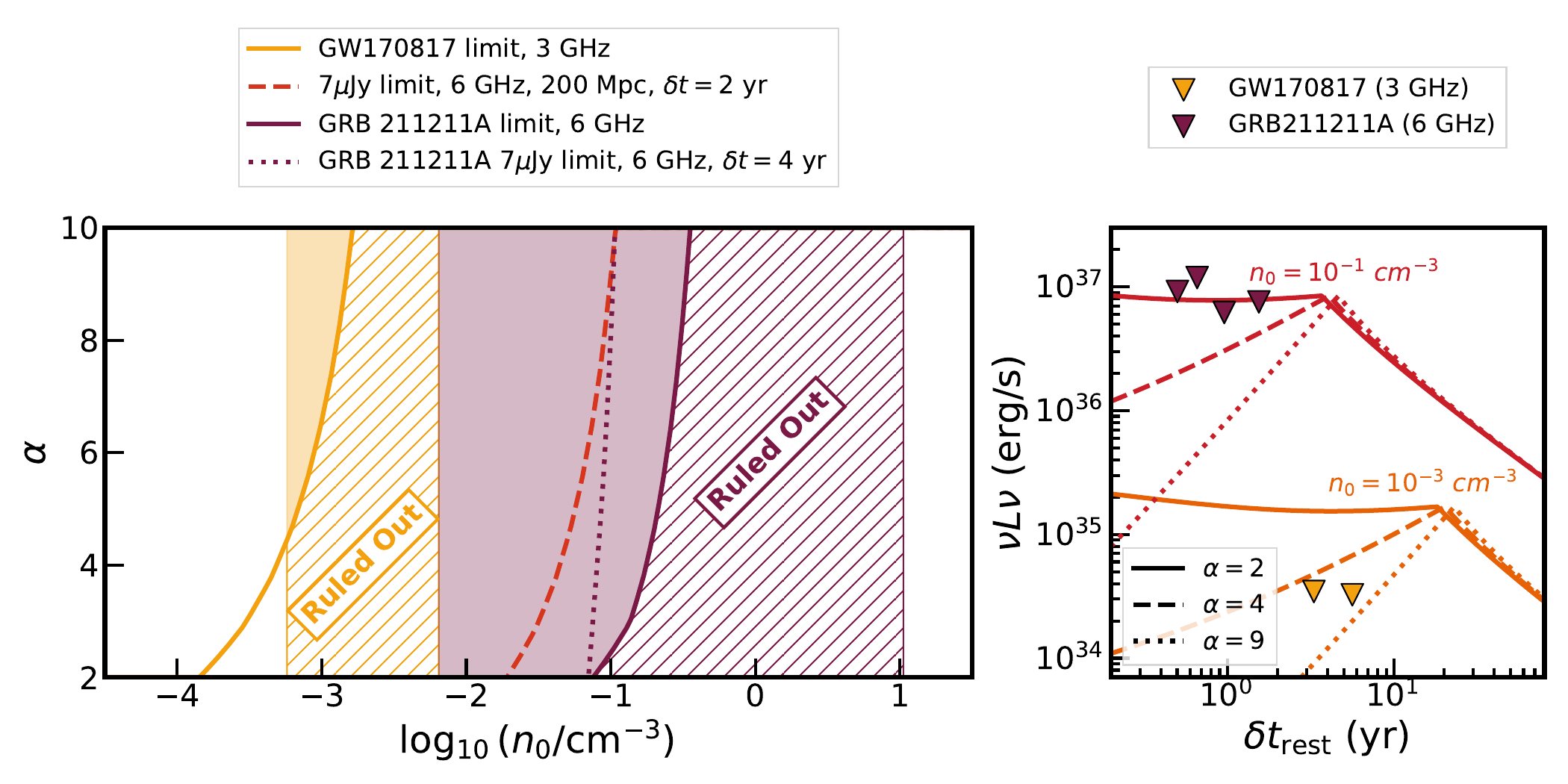}
    \caption{\emph{Left:} The $\alpha$ vs. $\log_{10}(n_{0}/{\rm cm}^{-3})$ parameter space. This space is created by generating light curves for varying $\alpha$--$n_0$ pairs, while holding $E_{\rm ej} = 10^{51}~{\rm erg}$, $M_{\rm ej} = 0.01~M_\odot$, $p = 2.2$, $\epsilon_{\rm e} = 0.1$, $\epsilon_{\rm B} = 0.01$ fixed, and finding the flux density of the light curve at the time of the radio observations.
    Solid lines indicate the limits placed on this space for GW\,170817 (yellow) and GRB\,211211A (purple) based on current radio limits. The orange dashed line represents a merger at $\sim 200~$Mpc, assuming an observation at $\delta t = 2~$years (similar to our final observation of GRB\,211211A at $\sim 1.7~$years) with a $3\sigma$ limit of $\lesssim 7~\mu$Jy at 6 GHz (comparable to the non-detections of 211211A and the sensitivity of the VLA for a 2 hour observation). The purple dotted line represents the limits that can be placed for an observation of 211211A at $\delta t =4~$years with a $3\sigma$ limit of $\lesssim 7~\mu$Jy at 6 GHz. Colored regions indicate the measured $n_{0}$ for 170817A (yellow) and 211211A (purple), where solid shaded regions are consistent with the current radio limits and hatched regions are ruled out by the current radio limits. \emph{Right:} Light curves for the radio non-detections of GW\,170817 and GRB\,211211A (yellow and purple triangles, respectively). Also plotted are kilonova afterglow models for $\alpha = 2,4,9$ (solid, dashed, and dotted, respectively) and two circumburst densities ($10^{-3}~{\rm cm}^{-3}$ in orange and $10^{-1}~{\rm cm}^{-3}$ in red), roughly corresponding to the densities allowed by the $\alpha$-$n_0$ parameter space for GW\,170817 and GRB\,211211A.}
    \label{fig:alpha_vs_n0}
\end{figure*}

While in our above analysis we assumed a single velocity shell of kilonova ejecta, in reality it is likely that the kilonova received a distribution of velocities, as described in Section~\ref{sec:Methods}. As a result, we can utilize our radio observations to place constraints on what the velocity distribution may be for GRB\,211211A. Continued follow-up of GW\,170817 at X-ray and radio wavelengths has led to constraints on the presence of a kilonova afterglow, indicating $\alpha \gtrsim 5$ (e.g. \citealt{2019ApJ...886L..17H, 2022ApJ...927L..17H}, though see also \citealt{2022MNRAS.510.1902T, 2022MNRAS.512.5572R, 2025MNRAS.539.2654K}).  

Given the different source of the shock powering the kilonova compared to the afterglow, it is not guaranteed that the kilonova will have similar microphysical values of $p$, $\epsilon_{\rm e}$, and $\epsilon_{\rm B}$ as the afterglow. However, as the kilonova ejecta travel through the same environment as the afterglow, it is likely that $n_0$ will remain the same between the two shocks. As a result, we focus on the constraints we can place on the $\alpha$--$n_0$ parameter space, since we have a constraint on $n_0$ from the afterglow. We generate a grid of light curves for various $\alpha$--$n_0$ pairs, where $\alpha = 2$--$6$ and $n_0 = 10^{-6}$--$10^{2}~{\rm cm}^{-3}$.
We fix $E_{\rm ej} = 10^{51}~$erg and $M_{\rm ej} = 0.01~M_\odot$, to be consistent with previous modeling of GW\,170817, which assumed $\beta \approx 0.3$--$0.35~M_\odot$. \citep{2019ApJ...886L..17H, 2019MNRAS.487.3914K, 2022ApJ...927L..17H}. We additionally fix $p = 2.2$, $\epsilon_{\rm e} = 0.1$, and $\epsilon_{\rm B} = 0.01$, consistent with our single velocity shell fiducial parameters (Table~\ref{tab:model}).

We generate a grid of light curves fixed at the redshift of GRB\,211211A ($z = 0.0763$) and matched to the deepest radio limits ($\nu = 6~$GHz). We then determine the flux density of the light curve at the time of the last two 211211A limits ($\delta t \approx 378$--$608~$days). For low values of $\alpha \lesssim 3$, the rise of the kilonova afterglow is shallow and as a result the emission may be detectable earlier with deeper limits, whereas for high values of $\alpha \gtrsim 3$ the rise is steep and therefore late time observations are more constraining (e.g. Figure~\ref{fig:alpha_vs_n0}, right). As a result, since the limit at $\delta t \approx 378~$days is deepest, this observation becomes more constraining for values of $\alpha \lesssim 3$ than the limit at $\delta t \approx 608~$days, and we must utilize both limits to properly explore the constraints on the $\alpha$-$n_0$ parameter space.
We determine the corresponding solution of both limits for $\delta t \approx 378$--$608~$days, and use the maximum (corresponding to the more constraining) value of $\alpha$ for each solution for the corresponding $n_0$. The solution is displayed in Figure~\ref{fig:alpha_vs_n0} (left). We are able to place meaningful constraints on $\alpha \gtrsim 2$ assuming $n_{0} \gtrsim 7.5\times 10^{-2}~{\rm cm}^{-3}$. 

Afterglow modeling of GRB\,211211A provides an independent constraint on the circumburst density, which has been measured to be $n_{0}\approx (0.65$-$1060) \times 10^{-2}~{\rm cm}^{-3}$ \citep[68\% confidence interval, ][]{2022Natur.612..223R}. Our constraints on the $\alpha$--$n_0$ space fall within the afterglow derived range, demonstrating that the VLA non-detections significantly constrain the presence of a kilonova afterglow.
We note that an additional VLA observation at $\delta t \approx 4~{\rm yr}$ after the burst to a $3 \sigma$ limit of $\lesssim 7~\mu{\rm Jy}$ would constrain a kilonova afterglow for {\it any} value of $\alpha$ for $n_0 \gtrsim 0.1~{\rm cm}^{-3}$ (Figure~\ref{fig:alpha_vs_n0}).

As a comparison, we next perform the same exercise for GW\,170817.
Several late time radio observations of GW\,170817 have led to deep constraints on the presence of a kilonova afterglow \citep{2019ApJ...886L..17H, 2022ApJ...927L..17H, 2022ApJ...938...12B, 2024GCN.36390....1E}. To match the GW\,170817 redshift and deepest radio limits, we generate a grid of light curves at $z = 0.0099$ and $\nu = 3~$GHz. We then find the flux density of the light curve at the time of the deepest GW\,170817 limit ($F_\nu \lesssim 4.8~\mu$Jy at $\delta t \approx 5.7~$yr, \citealt{2024GCN.36390....1E}). We determine the corresponding solution of this parameter space, as measured by the observed radio limits. The solution is displayed in Figure~\ref{fig:alpha_vs_n0}. For our chosen $E_{\rm ej}$, $M_{\rm ej}$, $p$, $\epsilon_{\rm e}$, and $\epsilon_B$, we are able to place meaningful constraints of $\alpha \gtrsim 2$ assuming $n_{0} \gtrsim 1.4\times 10^{-4}~{\rm cm}^{-3}$, whereas the circumburst density of GW\,170817 is estimated to be $n_{0} \approx (5.8$--$65) \times 10^{-4}~{\rm cm}^{-3}$ \citep[68\% confidence interval, ][]{2019ApJ...886L..17H}.
Within the measured $n_0$ range for GW\,170817, the non-detections require $\alpha \gtrsim 4$, similar to previous studies \citep{2019ApJ...886L..17H, 2022ApJ...927L..17H}. Importantly, while GRB\,211211A is a factor of $\approx 9$ further in distance than GW\,170817, the higher measured circumburst density for GRB\,211211A still allows us to place strong constraints on the $\alpha$--$n_0$ parameter space compared to GW\,170817. We note that this comparison assumes that all other parameters are the same across the two events. 

Finally, we investigate the constraints we can place on the $\alpha$--$n_0$ parameter space assuming a NS merger occurs at $z = 0.045$ ($\approx 200~$Mpc, the limits of LVK O4, \citealt{2020LRR....23....3A}). We generate a grid of light curves at this fixed distance following the same procedure detailed above for $\nu = 6~$GHz. We find the flux density of the light curve at $\delta t = 2~$yr after the merger (similar to our last observation of GRB\,211211A at $\sim 1.7~$yr), and determine the solution assuming a $3 \sigma$ limit of $\lesssim 7~\mu$Jy, the sensitivity reached by a 2 hour observation with the VLA at 6 GHz. Such an observation would lead to constraints on $\alpha$ assuming $n_{0} \gtrsim 1.9\times 10^{-2}~{\rm cm}^{-3}$. Given that the median measured $n_0$ for short GRBs is $\approx (0.3$--$1.5) \times 10^{-2}~{\rm cm}^{-3}$ \citep{2015ApJ...815..102F}, we would require such a merger at 200~Mpc to occur within a moderately high density environment in order to probe the kilonova afterglow scenario, as in GRB\,211211A. Later observations will place strong constraints on $\alpha \gtrsim 3$, similar to the simulated observation of GRB\,211211A at $\delta t \approx 4~$years.

We next explore an alternative explanation for the lack of kilonova afterglow emission. \citet{2020MNRAS.495.4981M} demonstrated that the kilonova afterglow emission could be suppressed if the kilonova ejecta has yet to catch up with the shell of the GRB. As the GRB jet plows through the ambient ISM, it will leave a cone of nearly empty space behind it. The kilonova ejecta will travel through this cone uninhibited, and without interacting with the ISM. As a result, prior to the time of collision ($t_{\rm col}$) of the kilonova ejecta and the GRB forward shock, the kilonova afterglow will be significantly fainter due to the lack of material with which to interact. Upon collision, a short-lived ($\sim~$yr) radio flare is predicted, followed by the standard kilonova afterglow. We can calculate $t_{\rm col}$ using Equation~9 from \citet{2020MNRAS.495.4981M}, where the parameters of interest are the energy of the GRB jet ($E_{\rm j}$), $n_0$, and the velocity of the kilonova ejecta. For the median afterglow parameters ($E_{\rm j} \approx 6 \times 10^{49}~$erg, $n_0 \approx 0.5~{\rm cm}^{-3}$), and assuming a kilonova velocity of $\sim 0.3 c$ \citep{2022Natur.612..223R}, we find $t_{\rm col} \approx 4.3~$years, whereas our final VLA observation of GRB\,211211A was at $\approx 1.7~$years. 
Nominally, this implies a suppression of the kilonova afterglow luminosity by a factor of $\gtrsim 100$ or more.
As a consequence, the constraints we placed on $\alpha$ may be weakened by this effect.
For the suppression estimated above and for large values of $\alpha$, the density constraint is similarly weakened by a factor of $\gtrsim 100^{4/(p+1)} \approx 300$ (i.e., a higher density would be allowed in Figure~\ref{fig:alpha_vs_n0}).
Finally, we note that in this scenario it is possible that future observations of GRB\,211211A may reveal a bright radio flare from the kilonova afterglow.

\section{Discussion}
\label{sec:Discussion}

\subsection{Implications for the Progenitor of GRB\,211211A}

While a compact object merger scenario remains the leading progenitor model of GRB\,211211A, the details of the compact objects involved remains unconfirmed.
Thus, we explore how our conclusions based on our analysis would change if GRB\,211211A was not produced by an NS-NS merger, for both the magnetar-boosted kilonova and the kilonova afterglow. 

If GRB\,211211A was instead produced by an NS-BH merger, whose timescales may also produce longer-lived $\gamma$-ray emission \citep[e.g.][]{2024MNRAS.527.7722S, 2024ApJ...963..112M, 2025MNRAS.540.2727L}, we would not expect to detect the magnetar-boosted kilonova signal, as the resulting BH remnant would not deposit any additional energy into the kilonova. This scenario is consistent with our observations, as we have not detected any such magnetar-boosted kilonova. If, however, GRB\,211211A was produced by an NS merger with a WD \citep[e.g.][]{2022Natur.612..232Y, 2023ApJ...947L..21Z, 2025ApJ...988L..46L}, a magnetar remnant could in principle still be produced; however, the rotational energy of the remnant may be considerably smaller than in a NS merger (e.g., \citealt{Margalit&Metzger17} predicted a final rotational energy of $\sim 10^{51}$ erg) and hence our constraints on $E_{\rm ej, max}$ are challenged to definitively rule out this scenario (Section~\ref{sec:Magnetar_Constraints}). Finally, if GRB\,211211A was produced by the accretion-induced collapse of a WD \citep[e.g.][]{Cheong+25,Combi+25}, the expected rotational energy of the resulting cooled magnetar remnant could deposit into the kilonova would be $\sim 3\times10^{52}~{\rm erg}$ \citep{2008MNRAS.385.1455M}. While we are unable to rule out this scenario based on our current limits placed on $E_{\rm ej, max}$ (Section~\ref{sec:Magnetar_Constraints}), future observations would be able to probe this scenario (Section~\ref{sec:future}).
Regardless of what type of compact objects merged  (e.g. NS-NS, WD-WD, NS-BH, or a WD-NS), our constraints on the presence of the kilonova afterglow remain valid if the red transient associated with GRB\,211211A is in fact produced by the radioactive decay of neutron rich ejecta produced during the merger (Section~\ref{sec:KN_afterglow}).

An alternative explanation for GRB\,211211A is the traditional long GRB progenitor of a collapsar, with unique properties. For example, \citet{2025ApJ...984...37W} found that GRB\,211211A could possibly be explained by a collapsar origin within a dust cloud, which in turn heats the dust and produces thermal emission, i.e. the red transient associated with GRB\,211211A. In this scenario, GRB\,211211A would also need to be at $z \approx 0.5$ to account for its lack of detected supernova counterpart. However we note that the apparent offset from the $z = 0.5$ galaxy and lack of underlying star formation is difficult to reconcile with this scenario \citep{2022Natur.612..223R}. Our radio observations are consistent with this progenitor model, as we would not expect to detect any late-time radio emission from such a scenario given the higher redshift. Additionally, the early radio observations of GRB\,211211A \citep{2022Natur.612..223R, 2022Natur.612..236M} were too shallow to rule out a radio supernova similar to SN 1998bw \citep{1998Natur.395..663K}.

\citet{2023ApJ...947...55B} also found that a collapsar progenitor scenario accompanied by an unusual supernova could explain the observed red transient following GRB\,211211A, at the assumed redshift of $z = 0.0763$ (see also: \citealt{2025arXiv250903003R}). In this scenario, the optical/infrared emission is still the result of $r$-process elements being produced, only the production of such ejecta occurs in the accretion disk around the collapsar central engine \citep{Siegel+19}, rather than mass ejected during a NS merger. \citet{2023ApJ...947...55B} found that there exists a narrow range of parameter space in which a collapsar could produce a transient that matches the observed behavior of GRB\,211211A.
They predicted that late-time radio observations of GRB\,211211A would distinguish between a collapsar and NS merger scenario that produced a magnetar, with the collapsar scenario producing radio emission that peaked on longer timescales ($\delta t_{\nu, \rm peak}\approx 5$--$15~$years) compared to the magnetar-boosted kilonova scenario ($\delta t_{\nu, \rm peak} \approx 1$--$5~$years), with peaks for both scenarios on the order of $F_{\nu, \rm peak} \approx 10$--$100~\mu$Jy. The values of $\delta t_{\nu, \rm peak}$ and $F_{\nu, \rm peak}$ are derived using \citet{np2011Natur.478...82N}, and therefore are similar, yet more simplistic, than our modeling framework. Overall, \citet{2023ApJ...947...55B} predict earlier $t_{\nu, \rm peak}$ and brighter $F_{\nu, \rm peak}$ for the magnetar-boosted kilonova than what we presented in Section~\ref{sec:Magnetar_Constraints}. 
As a result, we are able to fully rule out all of the magnetar-boosted kilonova models they present with our observations. However, our observations are unable to rule out the collapsar scenario presented in \citet{2023ApJ...947...55B}, and future observations will be needed to do so.

\subsection{The future of observing GRB\,211211A and other merger-driven long GRBs}
\label{sec:future}

\begin{figure}
    \centering
    \includegraphics[width = 0.48\textwidth]{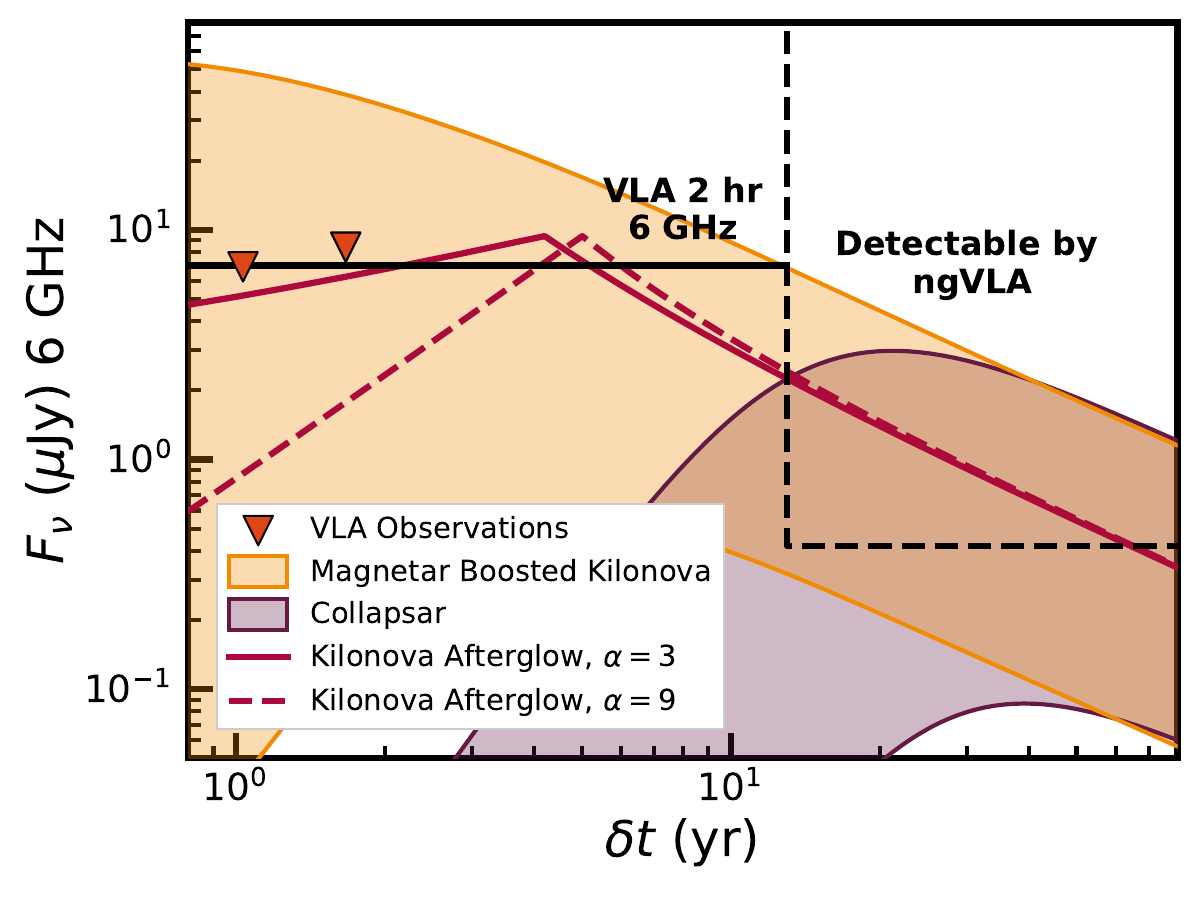}
    \caption{Expected radio light curves (assuming a singular bulk velocity) of two potential emission outcomes of 211211A: a collapsar scenario (purple, $E_{\rm ej} = 10^{52}$--$10^{53}~$erg, $M_{\rm ej} = 0.5$-$1.0~M_\odot$, e.g.  \citealt{2023ApJ...947...55B}) and a magnetar-boosted kilonova (orange, $E_{\rm ej} = 10^{52}$--$10^{53}~$erg, $M_{\rm ej} = 0.047~M_\odot$). Solid shaded regions indicate the range of light curves for the median afterglow parameters as measured by \citet{2022Natur.612..223R}.
    Also shown is two kilonova afterglow light curves (red, $E_{\rm ej} = 10^{51}~$erg, $M_{\rm ej} = 0.01 M_\odot$, e.g. \citealt{2019ApJ...886L..17H, 2022ApJ...927L..17H}) assuming stratified ejecta with velocity distribution indices of $\alpha = 3$ (solid) and $\alpha = 9$ (dashed) and $n_0 = 0.1~{\rm cm}^{-3}$ (See Section~\ref{sec:KN_afterglow} and \ref{sec:future}).
    Dark orange triangles represent the curent VLA upper limits for this burst. The solid black horizontal line indicates the $3\sigma$ sensitivity of the VLA for a 2 hour observation at 6~GHz. The vertical black dashed line indicate the expected time the ngVLA will come online, and the horizontal black dashed line indicates the expected $3\sigma$ sensitivity of the ngVLA at 8~GHz. We note that the SKA and DSA-2000 will also be able to probe the model light curves plotted here, however they will overall be less sensitive than the ngVLA.
    }
    \label{fig:FutureObs}
\end{figure}

We next explore the prospects for future observing campaigns of GRB\,211211A in the context of the magnetar-boosted kilonova, kilonova afterglow, and the unusual collapsar scenario presented in \citet{2023ApJ...947...55B}. We have already demonstrated that our VLA observations rule out a magnetar-boosted kilonova scenario down to $E_{\rm ej} \lesssim 4.4 \times 10^{52}~{\rm erg}$ using the afterglow derived parameters (Section~\ref{sec:Magnetar_Constraints}). Further observations with the VLA are unlikely to achieve much stronger constraints on this scenario, as the models bright enough to be detected by the VLA peak around the time of our observations (e.g. Figure~\ref{fig:LightCurves}). However, future planned radio facilities such as the next generation VLA (ngVLA\footnote{\url{https://ngvla.nrao.edu/}}), the Square Kilometer Array (SKA\footnote{\url{https://www.skao.int/en}}), and the Deep Synoptic Array 2000 (DSA-2000\footnote{\url{https://www.deepsynoptic.org/overview}}),  which should begin full science operations in the late 2020s--2030s, should be able to probe models down to $E_{\rm ej} \approx 2\times 10^{52}~{\rm erg}$ (Figure~\ref{fig:FutureObs}). Such future observations will not only place stronger constraints on the NS-NS merger scenario, but also other scenarios such as accretion-induced collapse of a WD \citep[e.g.][]{Cheong+25,Combi+25}. 

As we mentioned in Section~\ref{sec:KN_afterglow}, our current observations are able to rule out a KN afterglow for $n_0 \gtrsim 0.1~{\rm cm}^{-3}$, though the kilonova afterglow may be suppressed until $\delta t \approx 4.3~$years. For $n_0 = 0.1~{\rm cm}^{-3}$, the kilonova afterglow is also expected to peak around $\delta t \approx 4$--$5~$years (Figure~\ref{fig:FutureObs}), with a peak of $F_\nu \approx~ 9\mu$Jy, detectable with a 2~hour observation at 6~GHz with the VLA. Additionally, the fading kilonova afterglow should be detectable with the ngVLA/SKA/DSA-2000. If such radio emission is detected, this would provide strong support for an NS merger progenitor for GRB\,211211A.

We next utilize the collapsar model parameters presented in \citet{2023ApJ...947...55B} within our own modeling framework to explore what constraints can be placed with future observations. 
In the collapsar scenario ($E_{\rm ej} = 10^{52}$--$10^{53}~$erg, $M_{\rm ej} = 0.5$--$1.0~M_\odot$, e.g. \citealt{2023ApJ...947...55B}), we find $\delta t_{\rm peak} \approx 20$-$40~$years with $F_{\nu, \rm peak} \approx 0.1$--$3~\mu$Jy at 6~GHz (assuming the afterglow parameters\footnote{The fiducial parameters (Table~\ref{tab:model}) typically produce models similar or more optimistic to those presented in \citet{2023ApJ...947...55B}, and as a result we choose to focus on the afterglow parameters.} in Table~\ref{tab:model}, Figure~\ref{fig:FutureObs}).
With the current capabilities of the VLA, we would not be able to detect the radio emission. However, the ngVLA, SKA, and DSA-2000\footnote{The collapsar model at the frequency of the DSA-2000 (1.35~GHz) is $F_{\nu, \rm peak}\approx 0.2$--$6~\mu$Jy.} would be able to place constraints on the the collapsar scenario assuming the collapsar deposited  $E_{\rm ej} \gtrsim 3 \times 10^{52}~$erg into the ejecta. Importantly, the predicted late time behavior after the peak of this emission is indistinguishable from the magnetar-boosted kilonova emission. Therefore it is necessary to observe the rise of this emission, which will break the degeneracies between the magnetar-boosted kilonova and collapsar scenarios. As a result, the DSA-2000, which is expected to come online in the late 2020s \citep{2019BAAS...51g.255H}, will be crucial for constraining the rise of the emission.

Overall, with currently operational facilities such as the VLA, we predict that further observations of GRB\,211211A will only be able to place deeper constraints on the emergence of kilonova afterglow emission and the optimistic collapsar scnenario presented in \citet{2023ApJ...947...55B}. However, future planned facilities, such as the ngVLA, SKA, and DSA-2000, should be able to probe the peak and decline of the collapsar scenario, as well as the decline of the magnetar-boosted kilonova and kilonova afterglow scenarios to greater depths than currently possible.

In addition to GRB\,211211A, GRB\,060614 and GRB\,230307A are two other nearby ($z \lesssim 0.125$) long GRBs with red kilonova-like transients that may have been produced by a compact object merger \citep{2015NatCo...6.7323Y, 2024Natur.626..737L, 2024Natur.626..742Y,Cheong+25}. Previous radio observations of GRB\,060614 have led to some constraints on the magnetar-boosted scenario \citep{2016ApJ...819L..22H}, and given the age of GRB\,060614, it is unlikely that future observations will be much more constraining on the progenitor. However no such study has been done on the more recent GRB\,230307A. At a nearby distance of $\sim 300~$Mpc and with a relatively high circumburst density of $\sim 0.25~{\rm cm}^{-3}$ (\citealt{2024Natur.626..737L}, though see also \citealt{2024Natur.626..742Y}), GRB\,230307A represents an additional opportunity to place strong constraints on both the magnetar-boosted kilonova and kilonova afterglow scenarios. {Additionally, there is some evidence that features present in the $\gamma$-ray light curve, including the spectral lag and minimum variability timescale, may provide insight into the progenitor regardless of duration \citep{2022Natur.612..223R, 2022Natur.612..228T, 2022A&A...664A.177S, 2023NatAs...7...67G, 2023ApJ...954L...5V, 2026JHEAp..4900456M}, which will aid in the future identification of merger driven long GRBs.} The successful detection of a magnetar-boosted kilonova following a long GRB would have far reaching implications for the progenitors of these systems (helping to confirm an NS merger origin), the production of long GRBs from such systems, the NS EoS, heavy element production, and NS merger rates. Therefore, continued follow-up of this small but valuable population is paramount for constraining this emission. 

\section{Conclusions}
\label{sec:Conclusions}

We have presented new VLA radio limits of GRB\,211211A from $\approx 0.54$--$1.7~{\rm years}$. These limits represent some of the deepest limits placed following a GRB produced by an NS merger, due to the proximity of GRB\,211211A and its higher ISM density than GW\,170817. We used these limits, in addition to our updated light curve modeling, to place constraints on a magnetar-boosted kilonova, as well as a stratified velocity profile for an unboosted kilonova afterglow.
We have come to the following conclusions:

\begin{itemize}
    \item Assuming GRB\,211211A is the result of a NS merger in which a magnetar was produced, we are able to place limits on the maximum ejecta energy imparted on the kilonova of $E_{\rm ej, max} \lesssim 4.4 \times 10^{52}~{\rm erg}$ (using the median afterglow parameters). This rules out the presence of an indefinitely stable magnetar, but does not rule out a temporarily stable magnetar. If we instead assume fiducial microphysical parameters for the magnetar-boosted kilonova, we are able to rule out a temporarily stable magnetar ($E_{\rm ej, max} \lesssim 6.1 \times 10^{51}~{\rm erg}$).
    \item Using a sample of afterglow parameters derived by \citet{2022Natur.612..223R}, we are able to rule out an indefinitely stable magnetar ($E_{\rm ej, max} < 10^{53}~{\rm erg}$) for $\sim 66\%$ of the sample. However, we can only rule out a temporarily stable magnetar ($E_{\rm ej, max} < 10^{52}~{\rm erg}$) for $\sim 2\%$ of the sample.
    \item If the kilonova was unboosted (no magnetar formed), but acquired a stratified velocity profile, our limits are still able to place meaningful constraints on the presence of a kilonova afterglow, thanks to the high measured $n_0$ of GRB\,211211A.
    For an NS merger at $200~{\rm Mpc}$, deep VLA observations at 2 years post merger would be able to constrain the $\alpha$-$n_0$ space assuming $n_0 \gtrsim 1.9 \times 10^{-2}~{\rm cm}^{-3}$
    \item The radio emission from an unboosted kilonova afterglow may be suppressed until the kilonova ejecta catches up to the jet forward shock. In the case of GRB\,211211A, we predict this to occur around $\sim 4.3~{\rm years}$.
    \item While we are able to fully rule out the magnetar-boosted kilonva models presented in \citet{2023ApJ...947...55B}, our observations cannot rule out the collapsar scenario, though future observations may. We find that future observations with the VLA will not significantly increase the constraints we can place on the presence of a magnetar-boosted kilonova, but can probe the presence of a kilonova afterglow. Observations with future planned radio observatories should be able to probe the magnetar-boosted kilonova models and collapsar models to $\sim 2 $--$3 \times 10^{52}~{\rm erg}$, as well as the kilonova afterglow.
\end{itemize}

Going forward, additional radio campaigns of GRB\,211211A may reveal a radio flare due to an unboosted kilonova afterglow. While the proximity of GRB\,211211A presents one of the best opportunities to search for the signature of a magnetar-boosted kilonova, this study demonstrates the difficulty in constraining the presence of one based on the dependency of various model parameters. Future radio observations of nearby NS mergers, either discovered via GW observatories or $\gamma$-ray observatories, will be key to constraining both the magnetar-boosted kilonova scenario as well as the kilonova afterglow scenario, and will lend insight into the progenitors of these events.

\section{Acknowledgements}

G.S. acknowledges support for this work was provided by the NSF through a Student Observing Support award from the NRAO. B.M. acknowledges support by the National Science Foundation under grant number AST-2508620. B.D.M. acknowledges support from the National Science Foundation (grant AST-2406637) and the Simons Foundation (grant 727700).  The Flatiron Institute is supported by the Simons Foundation.
The Fong Group at Northwestern acknowledges the support of the National Science Foundation under grant No. AST-1909358 and CAREER grant No. AST2047919. W.F. gratefully acknowledges the support of the David and Lucile Packard Foundation, the Alfred P. Sloan Foundation, and the Research Corporation for Science
Advancement through Cottrell Scholar Award \#28284. E.B. acknowledges support from NSF and NASA grants. This study was enabled by a Radboud Excellence fellowship from Radboud University in Nijmegen, Netherlands. 
JCR was supported by NASA through the NASA Hubble Fellowship grant \#HST-HF2-51587.001-A awarded by the Space Telescope Science Institute, which is operated by the Association of Universities for Research in Astronomy, Inc., for NASA, under contract NAS5-26555.
GPL acknowledges support from the Royal Society (grant nos. DHF-R1-221175 and DHF-ERE-221005).
BPG acknowledges support from STFC grant No. ST/Y002253/1 and from The Leverhulme Trust grant No. RPG-2024-117.

The National Radio Astronomy Observatory is a facility of the National Science Foundation operated under cooperative agreement by Associated Universities, Inc. VLA observations for this study were obtained via projects VLA/22A-495 and VLA/23A-298. 

This research was supported in part through the computational resources and staff contributions provided for the Quest high performance computing facility at Northwestern University which is jointly supported by the Office of the Provost, the Office for Research, and Northwestern University Information Technology.

\facilities{VLA}
\software{CASA \citep{CASA}, pwkit \citep{2017ascl.soft04001W}, matplotlib \citep{matplotlib}, pandas \citep{pandas}, numpy \citep{numpy}}

\newpage 
\bibliographystyle{aasjournalv7}
\bibliography{library,journals_apj}

\clearpage

\end{document}